\documentclass[12pt,twoside]{article}
\usepackage[mathscr]{eucal}
\usepackage{amsmath,amsfonts,amssymb,amsthm,ulem,latexsym}

\usepackage{tabularx}
\usepackage{times}
\usepackage{cite}

\usepackage{multibox}

\usepackage{hyperref}

\def\be{\begin{equation}}
\def\ee{\end{equation}}
\def\ba{\begin{array}}
\def\ea{\end{array}}

\def\dps{\displaystyle}

\advance\textheight\topskip \textwidth 35.5pc \oddsidemargin 20pt
\renewcommand{\tilde}{\widetilde}
\renewcommand{\hat}{\widehat}

\newcommand{\assalgebra}{\mathscr}    

\newcommand{\binner}[2]{%
  {\langle}\kern-4.15pt{\langle}#1{,}\,#2{\rangle}\kern-4.15pt{\rangle}}

\newcommand{\ffrac}[2]{\raisebox{.5pt}%
  {\footnotesize$\displaystyle\frac{#1}{#2}$}\kern1pt}

\def\cN{\mathcal{N}}

\numberwithin{equation}{section} \makeatletter
\@addtoreset{equation}{section}

\makeatletter
\def\Appendix{\appendix
  \def\@seccntformat##1{Appendix~\csname the##1\endcsname.~~}}
\makeatother

\def\be{\begin{equation}}
\def\ee{\end{equation}}
\def\ba{\begin{array}}
\def\ea{\end{array}}

\def\dps{\displaystyle}

\def\ba{\begin{array}}
\def\ea{\end{array}}

\def\dps{\displaystyle}

\newcolumntype{x}[1]{%
>{\centering\hspace{0pt}}m{#1}}%
\newcolumntype{w}[1]{%
>{\raggedright\hspace{0pt}}m{#1}}%
\newcolumntype{z}[1]{%
>{\raggedleft\hspace{0pt}}m{#1}}%

\begin{document}

\begin{flushright}
FIAN-TD-2011-09 \\
\end{flushright}
\vspace{1.5cm}

\begin{center}

{\Large\textbf{Instantons and 2d Superconformal field theory}}

\vspace{1.4cm}

{\large{A.~Belavin$^{1}$, V.~Belavin$^{2}$, M.~Bershtein$^{1,3}$}}

\vspace{.7cm}

$^1$~\parbox[t]{0.85\textwidth}{\normalsize\it\raggedright
Landau Institute for Theoretical Physics, RAS, Chernogolovka, Russia}
$^2$~\parbox[t]{0.85\textwidth}{\normalsize\it\raggedright
Theoretical Department, Lebedev Physical Institute, RAS, Moscow, Russia}
$^3$~\parbox[t]{0.85\textwidth}{\normalsize\it\raggedright
Independent University of Moscow, Russia}

\vspace{1.5cm}

\begin{abstract}
A recently proposed correspondence between 4-dimensional $\cN=2$ SUSY $SU(k)$ gauge theories on $\mathbb{R}^4/\mathbb{Z}_m$ and $SU(k)$ Toda-like theories with $Z_m$ parafermionic symmetry is used to construct
four-point $\cN=1$ super Liouville conformal block, which corresponds to the particular case $k=m=2$.

The construction is based on the conjectural relation between moduli spaces of $SU(2)$ instantons on $\mathbb{R}^4/\mathbb{Z}_2$ and algebras like $\hat{gl}(2)_2\times \assalgebra{NSR}$. This conjecture is confirmed by checking the coincidence of number of fixed points on such instanton moduli space with given instanton number $N$ and dimension of subspace degree $N$ in the representation of such algebra.

\end{abstract}

\end{center}

\newpage


\section{Introduction}
Alday, Gaiotto and Tachikawa~\cite{Alday:2009aq} proposed correspondence between Liouville theory and four-dimensional $\cN=2$ supersymmetric gauge theories. This correspondence has been generalized for the conformal theories with additional symmetries, such as affine Lie algebras, $\assalgebra{W}$ algebras, parafermions \cite{Alday:2010vg,Wyllard:2009hg,MM,Taki,BF,Tachikawa:2011ParaL,Tanzini:2011ALE}.
In particular, it was suggested in \cite{BF} that the instanton calculus in the gauge theories on $\mathbb{R}^4/\mathbb{Z}_2$
give rise to the super-Virasoro conformal blocks. The idea to use the $\mathbb{Z}_2$ symmetric instanton moduli $\mathcal{M}_{\text{sym}}$ is based on its conjectural relation to the coset $\hat{gl}(n)_2/\hat{gl}(n-2)_2$ which is isomorphic to $\assalgebra{A}=\hat{gl}(2)_2\times \assalgebra{NSR}$ (see \cite{Nakajima1}-\cite{Feigin1} for the relation between instanton moduli spaces and algebras). This relation should mean that the algebra $\assalgebra{A}$ acts on the direct sum of certain cohomology of spaces $\mathcal{M}_{\text{sym}}$. In particular the algebra $\assalgebra{A}$ has a representation with basis labeled by fixed points for the torus action on the moduli spaces $\mathcal{M}_{\text{sym}}$. In section 3 we show that the number of fixed points actually coincide with the number of states in the certain representation of algebra $\assalgebra{A}=\hat{gl}(2)_2\times \assalgebra{NSR}$.

The $\mathbb{Z}_2$ symmetric instanton moduli $\mathcal{M}_{\text{sym}}$ decomposes into several connected components. The only two of these components are used in \cite{BF} as the new integration domain. The integral is interpreted the $\mathbb{Z}_2$ restricted instanton partition function of $SU(2)$ $\cN=2$ supersummetric pure gauge theory. This function coincides with so-called Whittaker or Gaiotto \cite{Gaiotto} limit of the four-point super-Liouville conformal block function for $\cN=1$ super Liouville theory (or equivalently norm of Whittaker vector). In section 3 we consider the number of fixed points on two used components. These numbers coincide with the number of states in the certain representation of algebra $\assalgebra{A}=\assalgebra{B}\times \assalgebra{B}\times \assalgebra{F} \times\assalgebra{NSR}$ where $\assalgebra{B}$ and $\assalgebra{F}$ denotes Heisenberg and Clifford algebras respectively.

Then we construct explicit expression of the general four-point super-Liouville conformal block function in terms of $\mathbb{Z}_2$ restricted instanton partition functions. For considered conformal block the matter fields in the fundamental representation of the gauge group should be included into the instanton calculations. Now, the integral over the moduli space involves zero modes of the matter fermions. After Nekrasov's deformation ~\cite{Nekrasov:2002qd}
(see also \cite{Flume:2002az,Tanzini:2002}) it takes efficient equivariant form and is handled be means of the localization method.  The instanton partition function coincide with the four-point conformal block up to additional factor related to the algebra $\hat{gl}(2)_2$. We found this factor and check the new representation up to the level $5/2$.

The paper is organized as follows. In Section \ref{ZmInstanton} we describe the structure of $\mathbb{Z}_2$ symmetric moduli spaces $\mathcal{M}_{\text{sym}}$.
The section~\ref{sec fixed points} is devoted to combinatorial study of fixed points on such space. The instanton moduli integral evaluation is the subject of Section \ref{Determinants}. We recall the definition of the four-point conformal block function in super-Liouville theory and give its new expression in terms of colored Young diagrams in Section \ref{4pointblock}.

\section{Modified moduli space}\label{ZmInstanton}

We begin by reminding ADHM construction \cite{adhm} of $N$-instanton solution in the case of $SU(2)$ gauge group
(see also \cite{BPSz,BZakh,aw,dm,DKM'}). ADHM date consist of complex
matrices, two $N\times N$ matrices $B_1$, $B_2$, a $N\times 2$
matrix $I$ and a $2\times N$ matrix $J$. The space $\mathcal{M}_N$ which defines all possible $N$-instanton solutions is given by the following set
of conditions:\\
i) Matrices   $B_1$, $B_2$, $I$ and $J$ satisfy the following equations
\begin{eqnarray}
\label{complexADHM}
&\left[ B_1, B_2\right]+IJ=0,
\end{eqnarray}
ii) The solutions related by $U(N)$ transformations
\begin{eqnarray}
B_i^{\prime}=gB_ig^{-1}, \, \, \, I^{\prime}=gI, \, \, \, J^{\prime}=Jg^{-1}; \, \, \, \, g\in GL(N)
\label{gauge}
\end{eqnarray}
are equivalent.\\
iii) Among vectors obtained by the repeated action of $B_1$ and $B_2$ on $I_{1,2}$ there exist
$N$ linear independent. Here $I_1$ and $I_2$ stand for the columns of  the matrix $I$ and
they are considered as two vectors of $N$-dimensional vector space $V$. The vector space $V$ is
attached to each point of $\mathcal{M}_N$. It formes a fiber of the $N$-dimensional
fiber bundle whose base is the moduli space $\mathcal{M}_N$ itself.

The subspace of the Moduli space $\mathcal{M}_{\text{sym}}$ is defined by the
following additional restriction of $\mathbb{Z}_2$ symmetry
\begin{equation}
B_{1}=-P B_{1} P^{-1};
B_{2}=-P B_{2} P^{-1};
\qquad I = P I;\qquad J=J P^{-1}.
\label{Zsym}
\end{equation}
where $P\in GL(N)$ is some gauge transformation. It is clear that $ P^2=1$. Hence the space $V$ decomposes $V_+ \oplus V_-$ where $Pv=v$ for $v \in V_+$ and $Pv=v$ for $v \in V_+$.

This new manifold $\mathcal{M}_{\text{sym}}$ is a disjoint union of
connected components $\mathcal{M}_{\text{sym}}(N_{+},N_{-})$,
where $N_{+}$ and $N_{-}$ denote dimensions of $V_{+}$ and $V_{-}$
correspondingly, $N_{+}+N_{-}=N$. These numbers are fixed inside given
connected component of $\mathcal{M}_{\text{sym}}$. Each component is connected and can be considered separately.

The construction of the instanton partition function involves the determinants  of the
vector field $v$ on $\mathcal{M}_N$, defined by
\begin{equation}
B_l\rightarrow t_l B_l ; \, \, \, \, I\rightarrow It_v; \, \, \,
\, J\rightarrow t_1 t_2 t_v^{-1}J,
\label{combinedaction}
\end{equation}
where parameters $t_l\equiv \exp \epsilon_l \tau$, $l=1,2$ and $t_v=\exp a \sigma_3 \tau$.

Fixed points, which are relevant for the determinants evaluation, are found from the
conditions:
\begin{equation}
t_l B_l=g^{-1}B_lg ; \, \, \, \, It_v=g^{-1}I; \, \, \, \, t_1 t_2
t_v^{-1}J=J g. \label{fixedpointconditions}
\end{equation}
The solutions of this system can be parameterized by pairs of Young diagrams $\vec{Y}=(Y_1,Y_2)$ such that the total number
of boxes $|Y_1|+|Y_2|=N$.
The cells $(i_1,j_1)\in Y_1$ and $(i_2,j_2)\in Y_2$ correspond  to the vectors
$B_1^{i_1} B_2^{j_1} I_1$ and  $B_1^{i_2} B_2^{j_2} I_2$ respectively.
It is convenient to use these vectors as a basis in the fiber $V$ attached to some fixed point.
Then the explicit form of the ADHM date for the given fixed point is defined straightforwardly
\begin{equation}
\begin{aligned}
&g_{ss'}=\delta_{ss'} t_1^{i_s-1}t_2^{j_s-1},\\
&(B_1)_{ss'}=d_{ss'}{}\delta_{i_s+1,i_{s'}} \delta_{j_s,j_{s'}},\\
&(B_2)_{ss'}=d_{ss'}\delta_{i_s,i_{s'}} \delta_{j_{s+1},j_{s'}},\\
&(I_\alpha)_{s}=\delta_{s,1_\alpha},\\
&J=0,
\end{aligned}
\end{equation}
where $s=(i_s,j_s), s'=(i_{s'},j_{s'})$ denotes the boxes of Young diagrams $Y_i$, $d_{ss'}=1$ if $s,s'$ belongs to the same Young diagram and $d_{s,s'}=0$ otherwise, $1_\alpha$ denotes the corner box ${(1,1)}$ of the diagram $Y_\alpha$.

Coming back to $\mathcal{M}_{\text{sym}}$ we note that it contains all fixed points of the vector field \eqref{combinedaction} found above. Eq.\eqref{Zsym}
defines the operator $P$ in the fixed point $\vec{Y}$
\begin{equation}
P(B_1^{i-1}B_2^{j-1} I_{\alpha})=(-1)^{i+j} B_1^{i-1}B_2^{j-1} I_{\alpha},
\end{equation}
so that the matrix elements can be found explicitly, $P_{ss'}=(-1)^{i_s+j_s}\delta_{ss'}$
In particular it follows that all fixed points belong to $\mathcal{M}_{\text{sym}}$.
The parity characteristic $P(s)=(-1)^{i_s+j_s}$ is assigned to each box in the Young diagrams
related to the fixed point. We adopt convenient notation from \cite{BF} that a box with coordinates of the same and different
parities are respectively white and black. Then $P(s)=1$ for the white boxes and $P(s)=-1$ for the black ones.
Therefore the fixed points can be classified by the numbers of white and black boxes, $N_{+}$ and $N_{-}$.
These numbers are the same as defined above, {\it i.e.} they are equal to the dimensions of the subspaces $V_{+}$ and $V_{-}$ of the fibers attached to those points
of $\mathcal{M}_{\text{sym}}$ which belong to the same component as the fixed point itself.

\section{Modified moduli space and $\hat{gl}(2)_2\times \assalgebra{NSR}$ algebra}\label{sec fixed points}

The norm of the Whittaker vector found in~\cite{BF} is equal to the sum of contributions of fixed points on connected components $\mathcal{M}_{\text{sym}}(N,N)$ and $\mathcal{M}_{\text{sym}}(N,N-1)$. In this section we
calculate the number of fixed points on such components and discuss the result from
the $\hat{gl}(2)_2\times \assalgebra{NSR}$ point of view.

It is convenient to introduce the generating function
\begin{equation}
\chi(q)=\sum\limits_{N}|\mathcal{M}_{\text{sym}}(N,N)|q^N+ \sum\limits_{N}|\mathcal{M}_{\text{sym}}(N,N-1)|q^{N-1/2},
\end{equation}
where $|\mathcal{M}_{\text{sym}}(N_+,N_-)|$ is a number of fixed points on $\mathcal{M}_{\text{sym}}(N_+,N_-)$. This number equals to the number of pairs of Young diagrams with $N_+$ white boxes and $N_-$ black boxes.

Denote by $d(Y)=N_{+}(Y)-N_{-}(Y)$ the difference between number of white and black boxes in Young diagram $Y$. For any integer $k$ we denote by
\begin{equation}\chi_k(q) = \sum\limits_{d(Y)=k} q^{\frac{|Y|}{2}},\end{equation}
the generating function of Young diagrams of given difference $d(Y)$. This function has the form:
\begin{equation} \chi_k(q)= q^{\frac{2k^2-k}2}\prod\limits_{m \ge 0} \frac1{(1-q^{m+1})^2}.\label{eq F_k}\end{equation}
The formula~\eqref{eq F_k} was proved in \cite[Sec. 5]{Kuznetsov} for $k=0$.  The factor $q^{\frac{2k^2-k}2}$ corresponds to the smallest
Young diagram with $d(Y)=k$. For $k>0$ this diagram consist of $2k-1$ rows of length
$2k-1,2k-2,\ldots,1$. For $k<0$ this diagram consist of $2|k|$ rows of length
$2|k|,2|k|-1,\ldots,1$.

The generating function of pairs Young diagrams with $N_+-N_-=k$ reads
\begin{equation}\chi^{(2)}_k = \sum\limits_{k_1+k_2=k} \chi_{k_1} \chi_{k_2},\end{equation}
Using~\eqref{eq F_k} and Jacobi triple product identity
\begin{equation}
\sum\limits_{n \in \mathbb{Z}} (-1)^nt^nq^{n^2}=\prod\limits_{m \ge 0} (1-q^{2m+2})(1-q^{2m+1}t)(1-q^{2m+1}t^{-1})
\end{equation}
we get
\begin{equation}
\chi(q)=\chi^{(2)}_0(q)+\chi^{(2)}_1(q)=
\prod\limits_{m \ge 0} \frac{(1-q^{m+\frac12})^2}{(1-q^{m+1})^3} = \chi_B(q)^3 \chi_F(q)^2 \label{eqchi},
\end{equation}
where
\begin{equation}
\chi_B(q)=\prod_{n\in\mathbb{Z},\, n>0} \frac{1}{(1-q^n)}
\end{equation}
\begin{equation}
\chi_F(q)=\prod_{r\in\mathbb{Z}+\frac12,\, r>0}(1+q^r).
\end{equation}
The first terms of the series for $\chi(q)$ looks as follows
\begin{equation}
\chi(q)=1+2q^{1/2}+4q+8q^{3/2}+16 q^2+28q^{5/2}+...
\end{equation}

The formula~\eqref{eqchi} looks very suggestive. The $\chi_B(q) \chi_F(q)$ equals to the character of standard representation of the $\assalgebra{NSR}$ algebra with generators $L_n$, $G_r$. The remaining part is related to the algebra $\assalgebra{B} \times \assalgebra{B} \times \assalgebra{F}$ where $\assalgebra{B}$ is the Heisenberg algebra with generators $b_n$ and relations $[b_n,b_m]=n\delta_{n+m}$ and $\assalgebra{F}$ is the Clifford algebra with generators $f_r$ and relations $\{f_r,f_s\}=r\delta_{r+s}$. 

Thus equation \eqref{eqchi} means that the generating function of numbers of fixed points on components $\mathcal{M}_{\text{sym}}(N,N)$ and $\mathcal{M}_{\text{sym}}(N,N-1)$ equals to the character of representation of the algebra $\assalgebra{A}=\assalgebra{B}\times \assalgebra{B}\times \assalgebra{F} \times\assalgebra{NSR}$. This representation theory point of view can be exploit similar to \cite{ALTF} (see also \cite{BB}).


One can consider the whole space $\mathcal{M}_{\text{sym}}$. The the generating function has the form
\begin{equation} \chi(q)=\sum\limits_{N}|\mathcal{M}_{\text{sym}}(N)|q^{\frac{N}2} = \prod_{n\in\mathbb{Z},\, n>0} \frac{1}{\left(1-q^{\frac{n}2}\right)^2}
\end{equation}
The result equals to the character of the certain representation of $\hat{gl}(2)_2\times \assalgebra{NSR}$ namely the tensor product of Fock representation of Heisenberg algebra, vacuum representation\footnote{more precisely character of vacuum representation in principal grading} of $\hat{sl}(2)_2$ and $NS$ representation of $\assalgebra{NSR}$. In other words the generating function of numbers of fixed points on $\mathcal{M}_{\text{sym}}(N)$ equals to the character of representation of the algebra $\assalgebra{A}=\hat{gl}(2)_2\times \assalgebra{NSR}$. 

Note that appearance of algebras $\assalgebra{B}\times \assalgebra{B}\times \assalgebra{F} \times\assalgebra{NSR}$ and $\hat{gl}(2)_2\times \assalgebra{NSR}$ may be related to the fact that $\hat{sl}(2)$ representation of level 2 can be realized by one bosonic and one fermionic field \cite{Fateev:1985mm}.

\section{Determinants of the vector field}\label{Determinants}
In~\cite{Nekrasov:2002qd, Flume:2002az} the form of $SU(k)$ $\cN=2$ supersummetric instanton partition function (in what follows we are dealing with $SU(2)$ case) was derived
as an integral of  the equivariant form. This form is defined in terms of the
vector field $v$ acting on the moduli space $\mathcal{M}_N$.

By means of the localization technique the evaluation of the
moduli integral is reduced to the calculation of the determinants \cite{Nekrasov:2002qd,Flume:2002az,Nakajima} of the vector field $v$ in the vicinity of fixed points
\begin{eqnarray}
\mathcal{Z}_N \left(a, \epsilon_1, \epsilon_2 \right)=  \sum_{n} \frac{1}{\det_n v}. \label{localization}
\end{eqnarray}
Here $n$ numerates fixed points of the vector field.

To evaluate the determinant of the vector field one needs to find all eigenvectors
of the vector field on the tangent space passing through the fixed points
\begin{equation}
\begin{aligned}
&t_i \delta B_i=\Lambda\, g \delta B_i g^{-1},\\
&\delta I t =\Lambda\, g \delta I,\\
&t_1 t_2 t^{-1} \delta J=\Lambda\, \delta J g^{-1}.
\end{aligned}
\end{equation}
This is equivalent to the following set of equations
\begin{equation}
\begin{aligned}
&\lambda\, (\delta B_i)_{s s'}=(\epsilon_i+\phi_{s'}-\phi_s)\, (\delta B_i)_{ss'},\\
&\lambda\, (\delta I)_{s p} =(a_p-\phi_s)\,  (\delta I)_{s p},\\
&\lambda\, (\delta J)_{p s}=(\epsilon_1+\epsilon_2-a_p+\phi_s)\, (\delta J)_{p s},
\label{linsyst}
\end{aligned}
\end{equation}
where $\Lambda=\exp \lambda\tau$, $g_{ss}=\exp \phi_s \tau$ and
\begin{equation}
\phi_s=(i_s-1)\epsilon_1+(j_s-1)\epsilon_2+a_{p(s)}.
\end{equation}
System \eqref{linsyst} gives all possible eigenvectors of the vector field. We should keep only those
which belong to the tangent space. Essentially this means excluding variations breaking
ADHM constraints. On the Moduli space
\begin{equation}
\label{complexvar}
\left[\delta B_1, B_2\right]+[B_1,\delta B_2]+\delta I J+ I \delta J=0.
\end{equation}
Gauge symmetry can be taken into account in the following manner.
We fix a gauge in which $\delta B_{1,2}, \delta I,\delta J$ are orthogonal to any gauge transformation of $B_{1,2},I,J$.
This gives additional constraint
\begin{equation}
\label{realgaugevar}
\left[\delta B_l,B_l^{\dagger}\right]+\delta I I^{\dagger}- J^{\dagger} \delta J=0.
\end{equation}
The variations in the LHS of \eqref{complexvar} and \eqref{realgaugevar} should be excluded
from \eqref{linsyst}. The corresponding eigenvalues are defined from the equations
\begin{equation}
\begin{aligned}
&t_1 t_2 (\left[\delta B_1, B_2\right]+[B_1,\delta B_2]+\delta I J+ I \delta J)=
\Lambda\, g \bigg(\left[\delta B_1,B_2\right]+[B_1,\delta B_2]+\delta I J+ I \delta J\bigg) g^{-1},\\
&\left[\delta B_l,B_l^{\dagger}\right]+\delta I I^{\dagger}- J^{\dagger} \delta J=\Lambda\,g\bigg(\left[\delta B_l,B_l^{\dagger}\right]+\delta I I^{\dagger}- J^{\dagger} \delta J\bigg)g^{-1}.
\end{aligned}
\end{equation}
One finds the following eigenvalues, which should be excluded from~\eqref{linsyst}:
\begin{equation}
\begin{aligned}
&\lambda=(\epsilon_1+\epsilon_2+\phi_s-\phi_{s'}),\\
&\lambda=(\phi_s-\phi_{s'}).
\end{aligned}
\end{equation}
Thus, the determinant of the vector field~\eqref{combinedaction} is given by
\begin{equation}
\det v=\frac{\prod_{s,s'\in \vec{Y}}(\epsilon_1+\phi_{s'}-\phi_s)(\epsilon_2+\phi_{s'}-\phi_s)
\prod_{l=1,2; s  \in \vec{Y}}(a_l-\phi_{s})(\epsilon_1+\epsilon_2- a_l+\phi_{s})}
{\prod_{s,s' \in \vec{Y}}(\phi_{s'}-\phi_s)(\epsilon_1+\epsilon_2-\phi_{s'}+\phi_s)}
\end{equation}

Now we consider the action of the vector field~\eqref{combinedaction} in $\mathcal{M}_{\text{sym}}$. The tangent
space is reduced by the additional requirement~\eqref{Zsym}
\begin{equation}
-\delta B_{1,2}=P \delta B_{1,2} P^{-1}; \qquad \delta I = P \delta I;\qquad \delta J=\delta J P^{-1},
\label{Zsymvar}
\end{equation}
or, on the level of the matrix elements,
\begin{equation}
-(\delta B_{1,2})_{ss'}=P(s) (\delta B_{1,2})_{ss'} P(s');
\quad (\delta I)_{sp} = P(s) (\delta I)_{sp};\quad (\delta J)_{ps}=(\delta J)_{ps} P(s),
\label{Zsymvarmatrix}
\end{equation}
The first relation in \eqref{Zsymvarmatrix} means that only eigenvectors $(\delta B_{1,2})_{ss'}$
with the different colors of $s$ and $s'$  belong to $Z_{\text{sym}}$. Similarly,
the second one leaves $(\delta J)_{ps}$ only if $s$ is white. The variations, which should be excluded
\eqref{complexvar} and \eqref{realgaugevar} belong to $\mathcal{M}_{\text{sym}}$ only for the matrix elements
between the states of the same color.\\
 Thus, we get the new determinant of the
vector field \eqref{combinedaction}
\begin{equation}
\begin{aligned}
\det{}' v=
\frac{\prod_{\substack{s,s'\in \vec{Y}\\P(s)\neq P(s')}}(\epsilon_1+\phi_{s'}-\phi_s)(\epsilon_2+\phi_{s'}-\phi_s)
\prod_{\substack{\alpha=1,2; s  \in \vec{Y}\\P(s)=1}}(a_{\alpha}-\phi_{s})
(\epsilon_1+\epsilon_2- a_{\alpha}+\phi_{s})}
{\prod_{\substack{s,s' \in \vec{Y}\\P(s)=P(s')}}(\phi_{s'}-\phi_s)(\epsilon_1+\epsilon_2-\phi_{s'}+\phi_s)}
\end{aligned}\end{equation}
Re-expressed in terms of arm-length and leg-length this expression gives
\begin{equation}
\begin{aligned}
 \det{}' v=\prod_{\alpha,\beta=1}^{2}
    \prod_{s\in {}^{\diamondsuit}Y_{\alpha}(\beta)}E\bigl(a_{\alpha}-a_{\beta},Y_{\alpha},Y_{\beta}\bigl|s\bigr)
(Q-E\bigl(a_{\alpha}-a_{\beta},Y_{\alpha},Y_{\beta}\bigl|s\bigr)),
\label{determinantvec}
\end{aligned}
\end{equation}
here $E\bigl(a,Y_1,Y_2\bigl|s\bigr)$ are
defined as follows
\be
\label{YTE}
E\bigl(a,Y_1,Y_2\bigl|s\bigr)=
a+b(L_{\scriptscriptstyle{Y_1}}(s)+1)-b^{-1} A_{\scriptscriptstyle{Y_2}}(s)\;,
\ee
where $A_{\scriptscriptstyle{Y}}(s)$ and $L_{\scriptscriptstyle{Y}}(s)$
are respectively the arm-length and the leg-length for a cell $s$ in $Y$.
The region ${}^{\diamondsuit}Y_{\alpha}(\beta)$  is defined as
\begin{equation}
{}^{\diamondsuit}Y_{\alpha}(\beta)=
\bigl\{(i,j) \in Y_{\alpha} \bigl| P\bigl(k'_j(Y_\alpha)\bigl)\neq P\bigl(k_i(Y_\beta)\bigl)\bigl\},
\end{equation}
or, in other words, the boxes having different parity of the leg- and arm-factors.
Formula \eqref{determinantvec} is similar to the equation (3.7) in \cite{Poghossian:2004}. So the contribution of the vector multiplet reads
\begin{equation}
\begin{aligned}
Z^{\text{\sf{sym}}}_{\text{\sf{vec}}}(\vec{a},\vec{Y})= \frac1{\det{}' v}
\end{aligned}
\end{equation}

We consider the $\mathcal{N}=2$ $SU(2)$ theory with $4$ fundamental hypermultiplets with masses $\mu$. These hypermultiplets give some additional contribution because of appearance of the $N$ null-modes for each kind of fermions in fundamental representation of the gauge group.
The amplitudes $\psi$ of the null-modes can be considered as the
fiber $V$ attached to one of the fixed point $\vec Y$.
The eigenvalues of the vector field are defined from the equation
\begin{equation}
\lambda\, \psi_{s} =(\mu+\phi_s)\,  \psi_{s},
\end{equation}
The corresponding contribution of the fundamental hypermultiplets
with masses $\mu_i$ looks as follows
\begin{equation}
Z_{\text{\sf{f}}}(\mu_i,\vec{a},\vec{Y})=\prod_{i=1}^{4}\prod_{\alpha=1}^{2}
    \prod_{s\in Y_{\alpha}}\bigl(\phi(a_{\alpha},s)+\mu_i\bigr),
\end{equation}

Considering the case of $\mathcal{M}_{\text{sym}}$ we impose some  restrictions on the set of eigenvectors for the fundamental multiplets. Namely we will assume that
$\psi \in V_{+}$, if $N$-even and $\psi \in V_{+}$, if $N$-odd.

The above consideration suggests the following form of the contributions of the fundamental hyper multiplets
\begin{equation}
\begin{aligned}
&Z^{\text{\sf{sym(0)}}}_{\text{\sf{f}}}(\mu_i,\vec{a},\vec{Y})=\prod_{i=1}^{4}\prod_{\alpha=1}^{2}
    \prod_{s\in Y_{\alpha},s-\text{white}}\bigl(\phi(a_{\alpha},s)+\mu_i\bigr),\\
&Z^{\text{\sf{sym(1)}}}_{\text{\sf{f}}}(\mu_i,\vec{a},\vec{Y})=\prod_{i=1}^{4}\prod_{\alpha=1}^{2}
    \prod_{s\in Y_{\alpha},s-\text{black}}\bigl(\phi(a_{\alpha},s)+\mu_i\bigr),\\
\label{determinant}
\end{aligned}
\end{equation}
The first of these partition functions correspond to the case with
even number of instantons, the second one correspond to the case
with odd number of instantons.

\section{Four-point Super Liouville conformal block}\label{4pointblock}
Two-dimensional super conformal Liouville field theory arises in the context of super string theory
in non-critical dimensions of spacetime \cite{Polyakov}. The Lagrangian of the theory reads
\begin{equation}
\mathcal{L}_{\text{SLFT}}=\frac1{8\pi}\left(  \partial_{a}\phi\right)
^{2}+\frac1{2\pi}\left(  \psi\bar\partial\psi+\bar\psi\partial\bar\psi\right)
+2i\mu b^{2}\bar\psi\psi e^{b\phi}+2\pi b^{2}\mu^{2}e^{2b\phi}\;.
\label{SL}
\end{equation}
Here $\mu$ is the cosmological constant and parameter $b$ is related to the central charge $c$
of the super-Virasoro algebra
\begin{equation}
\begin{aligned}
c=1+2Q^{2}\label{cQ}\;,\,\,\,
Q=b+\frac 1 b\;.
\end{aligned}
\end{equation}
In this paper we are interested in the Neveu-Schwarz sector of the super-Virasoro algebra
\be
\label{NRS}
\ba{l}
\dps
[L_n, L_m] = (n-m) L_{n+m} + \frac{c}{8}(n^3-n)\delta_{n+m}\;,
\\
\dps
\{G_r, G_s\} = 2L_{r+s}+\frac{1}{2} c(r^2-\frac{1}{4})\delta_{r+s}\;,
\\
\dps
[L_n, G_r] = (\frac{1}{2}n-r)G_{n+r}\;.
\ea
\ee
where the subscripts $m,n$ are integer and $r,s$ half-integer.
The NS fields are classified in highest weight representations of super-Virasoro algebra.
One of the central problems in CFT is the computation of the correlation functions of the primary fields
represented by the highest weight vectors. We denote them $\Phi_{\Delta}$ and $\Psi_{\Delta}$.
The highest weight vector $\Phi_{\Delta}$  is annihilated by all positive-frequency
generators and has the conformal dimension $\Delta$ defined by $L_0 |\Delta\rangle = \Delta |\Delta\rangle$,
while $\Psi_{\Delta}\equiv G_{-1/2} \Phi_{\Delta}$. Together fields $\Phi_{\Delta}$ and  $\Psi_{\Delta}$
form primary super doublet. In what follows we use the standard parametrization of
the conformal dimensions
\begin{equation}
\Delta(\lambda)=\frac{Q^2}{8}-\frac{\lambda^2}{2}.
\end{equation}
In \cite{Hadasz:2006sb,Belavin:2007zz,Belavin:2007gz,Belavin:2007eq,Hadasz:2007nt}
four-point correlation functions of the primary fields in NS sector were constructed
by means of so-called elliptic and $c$-recursion procedures. By means of the bootstrap technique
\cite{Belavin:1984vu} they are expressed in terms of the basic conformal blocks and structure constants
of the operator algebra \cite{Rashkov,Poghosian}. In particular,
the  $4$-point correlation function of bosonic primaries $\Phi_i$ with
conformal weights $\Delta_i$ is given by
\be
\ba{c}
\dps
\langle \Phi_1(q) \Phi_2(0) \Phi_3(1) \Phi_4(\infty) \rangle  =(q\bar q)^{\Delta - \Delta_1 - \Delta_2} \sum_\Delta \Big( C^\Delta_{12}C^\Delta_{34} F_0(\Delta_i|\Delta|q)F_0(\Delta_i|\Delta|\bar q)
\\
\\
\dps
\hspace{6cm} + \tilde C^\Delta_{12}\tilde C^\Delta_{34} F_1(\Delta_i|\Delta|q)F_1(\Delta_i|\Delta|\bar q)\Big)\;.
\ea
\ee
Below we quote the results of \cite{Belavin:2007zz} for the first few coefficients in the series expansion of the superconformal blocks $F_{0,1}$
\begin{equation}
\begin{aligned}
\label{superblock}
&F_0(\Delta_i|\Delta|q) = \sum_{N=0,1,...} q^N F^{(N)}(\Delta_i|\Delta)\;,\\
&F_1(\Delta_i|\Delta|q) = \sum_{N=1/2,3/2,...} q^N F^{(N)}(\Delta_i|\Delta)\;,
\end{aligned}
\end{equation}
up to $N=2$:
{\allowdisplaybreaks
\begin{align}
F^{(0)}={}&1,
\label{F0}
\\
F^{(\frac 12)}={}&\frac{1}{2\Delta},
\label{F1}
\\
F^{(1)}={}&\frac{(\Delta+\Delta_1-\Delta_2)(\Delta+\Delta_3-\Delta_4)}
{2\Delta},
\\
F^{(\frac 32)}={}&\frac{(1+2\Delta+2\Delta_1-2\Delta_2)
(1+2\Delta+2\Delta_3-2\Delta_4)}{8\Delta(1+2\Delta)}\nonumber
\\*
&{}+\frac{4(\Delta_1-\Delta_2)(\Delta_3-\Delta_4)}
{(1+2\Delta)(c+2(-3+c)\Delta+4 \Delta^2)},
\\
F^{(2)}={}&\frac{(\Delta+\Delta_1-\Delta_2)(1+\Delta+\Delta_1-\Delta_2)
(\Delta+\Delta_3-\Delta_4)(1+\Delta+\Delta_3-\Delta_4)}
{4\Delta(1+2\Delta)}
\nonumber
\\*
&{}+\frac{(\Delta^2-3(\Delta_1-\Delta_2)^2+2\Delta(\Delta_1+\Delta_2))
(\Delta^2-3(\Delta_3-\Delta_4)^2+2\Delta(\Delta_3+\Delta_4))}
{2\Delta(3+2\Delta)(-3+3 c+16\Delta)}
\nonumber
\\*
&{}+\frac{(\Delta_1-2(\Delta_1-\Delta_2)^2+\Delta_2+\Delta(-1+2\Delta_1+2\Delta_2))
(\Delta_1,\Delta_2\to\Delta_3,\Delta_4)}
{(c+2(-3+c)\Delta+4\Delta^2)(3+4\Delta(2+\Delta))}.
\end{align}
}
On the basis of the results of the previous section we suggest the following representation
for the NS four-point conformal blocks \eqref{superblock}:
 \begin{equation}
\begin{aligned}
\label{N1-conjecture}
\sum_{N=0,1,\dots} q^N \sum_{\vec{Y},\substack{N_{+}(\vec{Y})=N\\N_{-}(\vec{Y})=N}}
Z^{\text{\sf{sym(0)}}}_{\text{\sf{f}}}(\mu_i,\vec{a},\vec{Y})
Z^{\text{\sf{sym}}}_{\text{\sf{vec}}}(\vec{a},\vec{Y})
\;\;\;\; &{=}
\;\;\;(1-q)^A F_0(\Delta(\lambda_i)|\Delta(a)|q)\;,\\
\sum_{N=\frac 12,\frac 32,\dots} q^N \sum_{\vec{Y},\substack{N_{+}(\vec{Y})=N+\frac 12\\N_{-}(\vec{Y})=N-\frac 12}}Z^{\text{\sf{sym(1)}}}_{\text{\sf{f}}}(\mu_i,\vec{a},\vec{Y}) Z^{\text{\sf{sym}}}_{\text{\sf{vec}}}(\vec{a},\vec{Y})
\;\;\; &{=}
\;\;\;\frac 12(1-q)^A F_1(\Delta(\lambda_i)|\Delta(a)|q)\;.
\end{aligned}
\end{equation}
The parameters of the conformal block functions are related to those of the instanton partition function
in the following manner
\begin{equation}
\begin{aligned}[centered]
\mu_1=\frac Q2-(\lambda_1+\lambda_2),\qquad \mu_2=\frac Q2-(\lambda_1-\lambda_2), \\
\mu_3=\frac Q2-(\lambda_3+\lambda_4),\qquad \mu_4=\frac Q2-(\lambda_3-\lambda_4), \\
\end{aligned}
\label{parameters-mu}
\end{equation}
and
\begin{equation}
A=\bigg(\frac Q2-\lambda_1\bigg)\bigg(\frac Q2-\lambda_3\bigg).
\label{parameterA}
\end{equation}

The formula~\eqref{N1-conjecture} is the main result of this paper. We have checked this formula up to the level $5/2$.

\section*{Acknowledgments}
We are pleased to acknowledge useful conversations with Kostya Alkalaev ,
Borya Feigin, Misha Lashkevich and Lesha Litvinov.
The work was held within the framework of the Federal
programs ``Scientific and Scientific-Pedagogical
Personnel of Innovational Russia'' on 2009-2013 (state contracts No. P1339 and No. 02.740.11.5165) and supported by the joint RFBR-CNRS
grant No. 09-02-93106.

\providecommand{\href}[2]{#2}\begingroup\raggedright\endgroup

\end{document}